\documentclass[aps,prb,twocolumn,amsmath,amssymb,superscriptaddress,floatfix]{revtex4}
\usepackage{graphicx}
\usepackage{bm}
\usepackage[usenames]{color}
\bibstyle{apsrev.bib}

\newcommand{\be}{\begin{equation}}
\newcommand{\ee}{\end{equation}}
\newcommand{\beqn}{\begin{eqnarray}}
\newcommand{\eeqn}{\end{eqnarray}}

\begin{document}

\title{Infinite disorder scaling of random quantum magnets in three and higher dimensions}
\author{Istv\'an A. Kov\'acs}
\email{ikovacs@szfki.hu}
\affiliation{Department of Physics, Lor\'and E\"otv\"os University, H-1117 Budapest,
P\'azm\'any P. s. 1/A, Hungary}
\affiliation{Research Institute for Solid State Physics and Optics,
H-1525 Budapest, P.O.Box 49, Hungary}
\author{Ferenc Igl\'oi}
\email{igloi@szfki.hu}
\affiliation{Research Institute for Solid State Physics and Optics,
H-1525 Budapest, P.O.Box 49, Hungary}
 \affiliation{Institute of Theoretical Physics,
Szeged University, H-6720 Szeged, Hungary}
\date{\today}

\begin{abstract}
Using a very efficient numerical algorithm of the strong disorder renormalization group method we have extended
the investigations about the critical behavior of the random transverse-field Ising model in three and four
dimensions, as well as for Erd\H os-R\'enyi random graphs, which represent infinite dimensional lattices.
In all studied cases an infinite disorder quantum critical point is identified, which ensures that
the applied method is asymptotically correct and the calculated critical exponents tend to the exact values
for large scales. We have found that the critical exponents are
independent of the form of (ferromagnetic) disorder and they vary smoothly with the dimensionality.
\end{abstract}


\maketitle
\section{Introduction}
Quantum phase transitions are among the fundamental problems of modern physics, the properties of
which are studied in solid state physics, quantum field-theory, quantum information and statistical mechanics\cite{sachdev}.
These transitions take place at $T=0$ temperature, i.e. in the ground state of the quantum system by varying a control
parameter, such as the strength of a transverse field. One basic question in this
field of research is how quenched disorder influences the properties of quantum phases and phase transitions. In this
respect quantum spin glasses and the glass transition are particularly interesting\cite{qsg}. This latter problem theoretically
is very challenging, since the corresponding quantum state is the result of an interplay between quantum and disorder fluctuations, strong correlations and frustration.

One of the paradigmatic models of random quantum magnets with a discrete symmetry is the random transverse-field
Ising model (RTIM), which is defined by the Hamiltonian:
\be
{\cal H} =
-\sum_{\langle ij \rangle} J_{ij}\sigma_i^x \sigma_{j}^x-\sum_{i} h_i \sigma_i^z\;.
\label{eq:H}
\ee
Here the $\sigma_i^{x,z}$ are Pauli-matrices and $i$, $j$ denote sites of a lattice (or a graph).
Experimentally the RTIM is closely related to the compound\cite{experiment} $\rm{LiHo}_x\rm{Y}_{1-x}\rm{F}_4$,
in which there is a dipole-coupling between the Ising
spins, thus the interaction is long-ranged. Applying a magnetic field $H_t$ transverse to the Ising axis results in a transverse field of strength, $h_i=H_t^2$, but this transverse field induces a random longitudinal field\cite{rf} via the
off-diagonal terms of the dipolar interaction.
In the theoretical investigations the interactions in the RTIM are
generally assumed to be short-ranged, thus the first sum in Eq.(\ref{eq:H}) runs over nearest neighbors.
Furthermore
the $J_{ij}$ couplings and the $h_i$ transverse fields are independent random numbers, which are taken from the
distributions, $p(J)$ and $q(h)$, respectively. For random ferromagnets we have $J>0$, whereas for spin-glasses
there are both ferro- and antiferromagnetic couplings. Here we are basically interested in the former problem.

Detailed theoretical results about the RTIM are known in one dimension (1D) due to a complete analytical
solution of a renormalization group (RG) treatment\cite{fisher}. The RG results are expected to be asymptotically exact
in the vicinity of the critical point (and also in the Griffiths-phase, as long as dynamical singularities are
concerned\cite{igloi02}), which is indeed demonstrated by a comparison with independent analytical\cite{mccoywu,
shankar} and numerical\cite{young_rieger96,bigpaper} works. One
important observation, that the critical properties of the 1D model are governed by an
infinite disorder fixed point (IDFP), in which the strength of disorder growths without limit during renormalization\cite{danielreview} and thus become dominant over quantum fluctuations.

The IDFP scenario is found to be valid for the 2D RTIM, too, as observed in numerical RG studies\cite{motrunich00,lin00,karevski01,lin07,yu07,ladder,2dRG} and
in Monte Carlo (MC) simulations\cite{pich98}. The calculated critical exponents are in agreement with the MC results
about the 2D random contact process\cite{vojta09}, which is a simple nonequilibrium model of spreading
infections. The $d$-dimensional random contact process is expected to be in the same universality
class\cite{hiv} as the RTIM, at least for strong enough disorder. 

In three dimensions, which is connected to real quantum magnets,
no quantitative results are known, so far. Analysis
of the numerical RG trajectories lead to the conclusion\cite{motrunich00}, that the critical behavior in this case
is probably controlled
by an IDFP, but no estimates about the critical exponents are available. For even higher dimensions it is
unclear, if the IDFP scenario stays valid for any finite value of $d$, or there is some upper critical dimension, $d^c$,
so that for $d>d^c$ the critical behavior is of conventional disorder type. We note that the large-$d$ limit of the
problem is qualitatively relevant for models with long-range interactions. In this respect the
critical behavior of the (non-random) $\rm{LiHo}\rm{F}_4$ system 
has been the subject of intensive MC simulations\cite{LiHo}.

In this paper we extend the investigations about the critical behavior of the RTIM into the hitherto unexplored
three and higher dimensions. Here we have developed a considerably improved numerical algorithm of the strong disorder
RG (SDRG) procedure and study large samples with $N \gtrsim 10^6$ sites. In 3D and 4D we consider hypercubic lattices
and we study the large-D limit of the problem, too. This latter
is realized by Erd\H os-R\'enyi (ER) random graphs\cite{erdos_renyi} consisting of $N$ sites and $kN/2$ edges ($k>2$),
which are at random positions. The sizes of the largest systems we studied are shown in Table \ref{table:1}.
We note that the numerical algorithm of the SDRG method is completely different from that used in the 2D
case, however, the steps used to calculate the critical parameters as well as 
the method of analysing the results are similar to that used in 2D\cite{2dRG}.

The structure of our paper is the following. The SDRG method and its improved algorithm used in this paper is
described in Sec.\ref{Sec:SDRG}. Results about the critical parameters are calculated in Sec.\ref{Sec:critical}
and discussed in Sec.\ref{Sec:discussion}.

\section{SDRG procedure}
\label{Sec:SDRG}

In the calculation we used the SDRG procedure\cite{im}, which has been introduced by Ma,
Dasgupta and Hu\cite{mdh}. In this method, which works for random ferromagnets, at each step of the renormalization the largest local term
in the Hamiltonan (either a coupling or a transverse field) is eliminated and new terms are generated between remaining sites by second-order perturbation method. After decimating a strong coupling, say $J_{ij}$, the two connected spins form a spin cluster having
an additive moment, $\tilde{\mu}=\mu_i+\mu_j$, which is placed in an
effective transverse field of strength: $\tilde{h}=\dfrac{h_i h_j}{J_{ij}}$. After decimating a large transverse field, $h_i$,
the actual spin is eliminated and new effective
couplings are generated between each pair of spins being nearest neighbors to the decimated site, say $j$ and $k$,
having a value: $\tilde{J}_{jk}=\dfrac{J_{ji}J_{ik}}{h_i}$. If at one step two parallel couplings appear between
two neighboring sites the maximum of them is taken.
Application of this ``maximum rule'' is exact at an IDFP and results in simplifications of the RG procedure.

\begin{figure}[h!]
\begin{center}
\includegraphics[width=3.4in,angle=0]{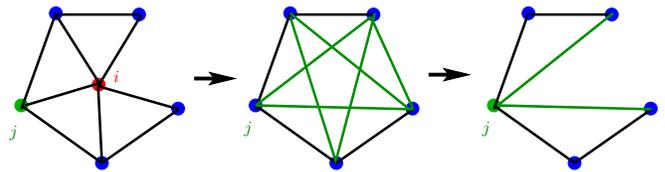}
\end{center}
\vskip -.5cm
\caption{
\label{fig_1} (Color online)
Illustration of the renormalization procedure. In the original cluster (left panel) the central site, $i$, is to be
decimated. In the na\"{\i}ve implementation (middle panel) several new couplings are
generated. In the improved algorithm (right panel) just the couplings to the relevant site, $j$, are created.
}
\end{figure}

Here we have developed an optimized algorithm, which needs $t \sim \mathcal{O}(N \log N + E)$ time to renormalize
a cluster with $N$ sites and $E$ edges up to the last spin,
irrespective of the dimension and topology of the cluster\cite{2D_performance}.
In this algorithm terms in the Hamiltonian are decimated in descending
order in energy and we have applied the following theorem for transverse field decimation.
According to this theorem for a decimated site, $i$,
there is always one relevant neighboring site, $j$, so that after decimating $i$ only those renormalized couplings should be
created, which connect
$j$ with its new neighboring sites\cite{selection}. All the couplings which start from other nearest neighbors of $i$
(and does not end at $j$) are irrelevant and need not be created. In this way during one RG step not only the number of sites is reduced (by one),
but the number of couplings, as well. Using this algorithm we avoid to generate almost fully connected clusters,
which is the main drawback of the na\"{\i}ve implementation of the method in higher dimensions\cite{2dRG}, having a performance: 
$t \sim \mathcal{O}(N^3)$. Renormalization with the na\"{\i}ve and the improved algorithms is illustrated in Fig.\ref{fig_1}.

\section{Calculation of critical parameters}
\label{Sec:critical}

In the actual calculation - in order to check universality and to control the disorder dependence
of the estimates - we have used two different
forms of randomness. Both have the same uniform distribution of the couplings:
$p(J)=\Theta(J)\Theta(1-J)$ ($\Theta(x)$ being the Heaviside step-function), which are ferromagnetic. For the 'box-$h$' disorder also the
transverse fields are uniformly distributed:
$q(h)=\dfrac{1}{h_b}\Theta(h)\Theta(h_b-h)$, 
whereas for the 'fixed-$h$' model we have a constant transverse field\cite{constant}: $h_i=h_f$. We used the logarithmic
variable, $\theta=\ln(h_b)$ or $\theta=\ln(h_f)$, as a quantum control parameter.
We have checked that the computational time to renormalize an $N=10^3$ ($N=10^6$) cluster is typically
$\sim 0.015$ ($\sim 50$) second (in a 2.4GHz processor), which does not depend on the dimension and the
topology of the cluster.
The numbers of realizations used in the calculations were typically $40000$ but even for the largest sizes we
have at least $10000$ samples.

In the first step of the calculation for each random sample, $\alpha$, we have determined a pseudo-critical point, $\theta_c(\alpha,N)$,
by a variant of the doubling method. In this procedure\cite{2dRG} we
glue together two identical copies ($\alpha,\alpha'$) of the
sample by surface couplings\cite{doubling_ER} and renormalize it up to the last site for different values of
the control parameter, $\theta$. The renormalization is found to be qualitatively different for $\theta<\theta_c(\alpha,N)$
and for $\theta>\theta_c(\alpha,N)$.
For weak quantum fluctuations, $\theta<\theta_c(\alpha,N)$, the last decimated spin cluster contains equivalent sites
of $\alpha$ and $\alpha'$. These sites and thus the two replicas are correlated and we call this cluster
as a \textit{correlation cluster}. The moment of the correlation cluster, $\mu(\alpha,N)$, goes to zero as the pseudo-critical
point is approached. On the contrary for $\theta>\theta_c(\alpha,N)$ in
the last decimated spin cluster there are no equivalent sites
of $\alpha$ and $\alpha'$ and thus there is no correlation cluster. 

\begin{figure}[h!]
\begin{center}
\includegraphics[width=3.4in,angle=0]{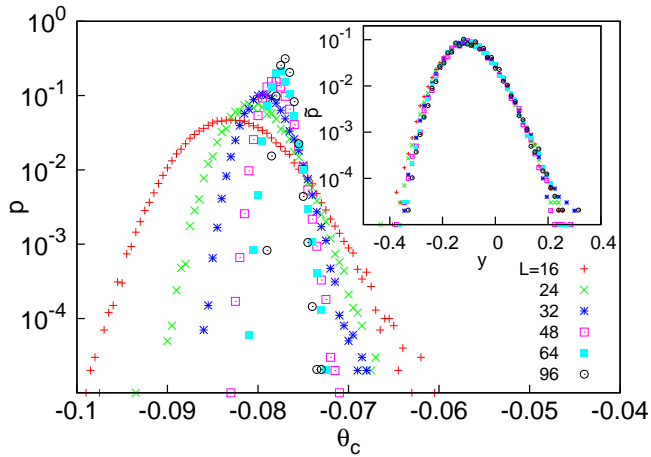}
\end{center}
\vskip -.5cm
\caption{
\label{fig_2} (Color online)
Distribution of the pseudo-critical points, $\theta_c(\alpha,N)$, for various sizes, $N=L^3$, for fixed-$h$ randomness
for the 3D model\cite{note_distr}.
In the inset the scaled distributions are shown as a function of
$y=(\theta_c(\alpha,N)-\theta_c)N^{1/d\nu}$, see the text.
}
\end{figure}

\begin{figure}[h!]
\begin{center}
\includegraphics[width=3.4in,angle=0]{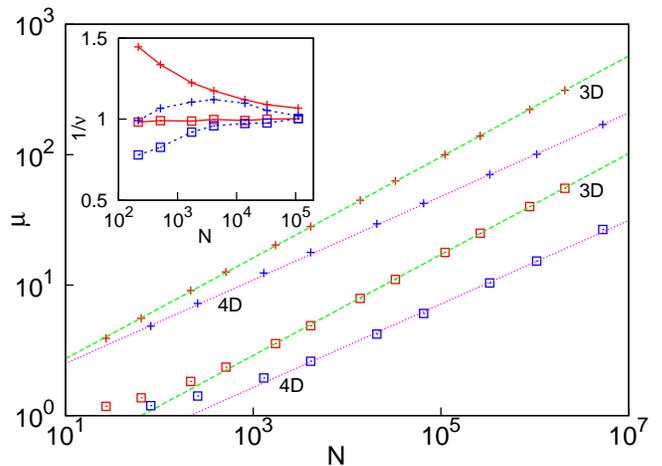}
\end{center}
\vskip -.5cm
\caption{
\label{fig_3} (Color online)
 Average moment of the correlation cluster at the critical point vs. the size of the system in a log-log plot
for the 3D and 4D models for two types of randomness (box-$h$: $\boxdot$, fixed-$h$: $+$). The slope of the straight lines is given by: $d_f/d=0.387$
and $d_f/d=0.32$ for 3D and 4D, respectively.
Inset: Finite-size effective exponents, $\nu_s$ (blue - dashed) and $\nu_w$ (red - full), for the 3D model for two types of randomness. In all cases the error of the calculation is smaller than the size of the symbol.
}
\end{figure}

In the second step we have studied the size-dependence of the distributions of the pseudo-critical points,
which is illustrated in Fig.\ref{fig_2} for the 3D model. From the
scaling of the width, $\Delta \theta_c(N) \sim N^{-1/d\nu_w}$, and from the scaling of the mean value:
$|\theta_c-\overline{\theta_c(N)}| \sim N^{-1/d\nu_s}$ we have obtained the critical exponents, $\nu_w$ and $\nu_s$,
respectively. We have calculated size-dependent effective exponents by two-point fits (comparing the results
for sizes $N$ and $N/2^d$), which are then
extrapolated. The effective exponents for the 3D model are shown in the inset of Fig.\ref{fig_3} for the two different randomnesses. As in this example we have generally observed that the extrapolated critical exponents are universal, i.e.
randomness independent. Estimates of the exponents are presented in Table \ref{table:1}, together with the values of the \textit{true} critical points, $\theta_c^{(b)}$ and $\theta_c^{(f)}$, for the two randomnesses, respectively. One
can notice in this Table that the error of the estimates is increasing with the dimensionality and the finite-size
corrections are considerably strong for 4D and for ER random graphs.

Having accurate estimates for the critical points we have renormalized the systems at $\theta_c$
and studied the scaling behavior of
the moment of the correlation cluster, $\mu(\alpha,N)$, as well as that of the log-energy parameter,
$\gamma(\alpha,N)=-\ln \epsilon(\alpha,N)$.
The average moment is found to scale as: $\overline{\mu}(N) \sim N^{d_f/d}$, where $d_f$ is the fractal dimension of the
correlation cluster. We illustrate this relation in Fig.\ref{fig_3} for the 3D and the 4D models,
in which $\overline{\mu}(N)$ is shown
as a function of $N$ in a log-log scale. Indeed, for not too small systems, $N>1000$, the points are very well on
straight lines, the slope of which being the same for the two different randomnesses for the same $d$. From the cluster moment
the magnetization is calculated as, $m=\overline{\mu}(N)/N$, thus we have the scaling relations: $x/d=\beta/(d\nu)=1-d_f/d$,
where $\beta$ is the magnetization exponent and $x=\beta/\nu$.
Estimates for the exponents $x/d$, which are calculated through two-point fits are shown in Table \ref{table:1}.

\begin{figure}[h!]
\begin{center}
\includegraphics[width=3.4in,angle=0]{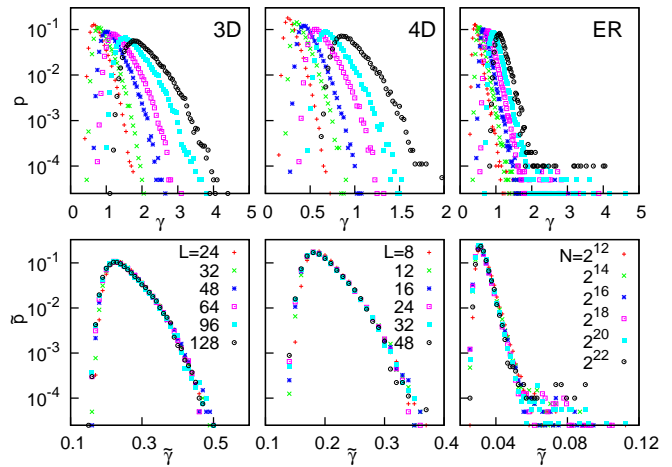}
\end{center}
\vskip -.5cm
\caption{
\label{fig_4} (Color online)
Distribution of the log-energy parameters at the critical point in 3D, 4D and in the
ER random graph for fixed-$h$ randomness and
for different sizes (upper panel)\cite{note_distr}.
In the lower panel the scaled distributions are shown, as described
in the text. The constants in the scaling forms are $\gamma_0=-0.33$ (3D), $\gamma_0=-0.23$ (4D) and $\gamma_0=-0.5$, $\ln N_0=-5. $ (ER).
}
\end{figure}

The energy parameter, $\epsilon(\alpha,N)$, is given by the value of the smallest effective transverse field,
not considering that of the correlation cluster (if any). 
The distribution of the log-energy parameter, $\gamma(\alpha,N)$, is shown
in the upper panel of Fig.\ref{fig_4}, for the different models. As a clear indication of infinite disorder scaling
the width of the distribution is increasing with $N$. In 3D and 4D the appropriate scaling variable is $\tilde{\gamma}=(\gamma(N)-\gamma_0)N^{-\psi/d}$ ($\gamma_0$ is a constant),
as illustrated with the data collapse in the lower panel of Fig.\ref{fig_4}. The critical exponent $\psi$
has been calculated from the optimal collapse of the distributions, as well as from two-point fits by
comparing the mean values
$\overline{\gamma}(N)$ and $\overline{\gamma}(N/2^d)$,
which are presented in Table \ref{table:1}. The ER random graphs are infinite dimensional objects and in this
case the broadening of the distribution of the log-energy parameter is found to scale with $\ln N$. As a good
scaling combination we have here $\tilde{\gamma}=(\gamma(N)-\gamma_0)(\ln N/N_0)^{-\omega}$ ($\gamma_0$ and $N_0$ are constants),
which is illustrated
with the data collapse in the lower panel of Fig.\ref{fig_4}, with an exponent $\omega=1.3(2)$. Thus the width
of the distribution increases somewhat faster than linear in $\ln N$, which fact justifies that also for ER
random graphs the critical behavior of the RTIM is controlled by a logarithmically infinite disorder fixed point.

\section{Discussion}
\label{Sec:discussion}

\begin{table}
\caption{Critical properties of the RTIM in three and four dimensions and in ER random graphs.
$N_{max}$ denotes the number of spins in the largest finite systems
used in the RG calculation. In the last four lines by estimating the different critical exponents
results obtained by the two forms of disorder are taken into account.
 \label{table:1}}
 \begin{tabular}{|c|c|c|c|c|c|}  \hline
    & 3D & 4D & ER \\ \hline
$N_{max}$   & $128^3$ & $48^4$ & $2^{22}$ \\ \hline
$\theta_c^{(b)}$    & $2.5305(10)$ & $3.110(5)$ & $2.775(2)$ \\ 
$\theta_c^{(f)}$    & $-0.07627(2)$ & $-0.04698(10)$ & $-0.093(1)$ \\ \hline
$d\nu_w$    & $2.90(15)$ & $3.30(15)$ & $7.(2)$ \\ 
$d\nu_s$  & $2.96(5)$  & $2.96(15)$ & $5.(1)$ \\ \hline
$x/d$    & $0.613(5)$ & $0.68(3)$ & $0.81(4)$ \\ \hline
$\psi$    & $0.46(2)$ & $0.46(2)$ & $-$ \\ \hline
  \end{tabular}
  \end{table}
Our numerical RG results indicate that the critical behavior of the random transverse-field Ising
model in three and four dimensions as well as in the ER random graph is controlled by infinite disorder
fixed points. This fact
justifies the use of the SDRG method and ensures that the calculated numerical results about the critical
exponents tend to be asymptotically correct for large sizes. Since the ER random graph represents the large-dimensional limit of the problem, infinite disorder scaling is expected to be valid at any dimensions. The critical exponents presented
in Table \ref{table:1} are found to be the same for the two types of ferromagnetic disorder used in our
numerical study. We expect therefore that the IDFP-s are attractive, at least for strong enough disorder.
(For the weak-disorder behavior of the systems we can not make any definite statement using the
SDRG method.) We note also that at the IDFP frustration does not matter, thus quantum spinglass
and random quantum ferromagnet has the same infinite disorder fixed point.

Singularities of the thermodynamic quantities at small temperatures involve the exponents in Table \ref{table:1}.
For example the susceptibility and the specific heat behave as: $\chi(T) \sim (\log T)^{(d-2x)/\psi}/T$ and
$C_V(T) \sim (\log T)^{d/\psi}$, respectively\cite{danielreview,im}. The analogous expressions for the ER random graph are:
$\log[T \chi(T)]\sim (1-2x/d)(\log T)^{1/\omega}$ and $\log [C_V(T)] \sim (\log T)^{1/\omega}$.

\begin{figure}[h!]
\begin{center}
\includegraphics[width=3.4in,angle=0]{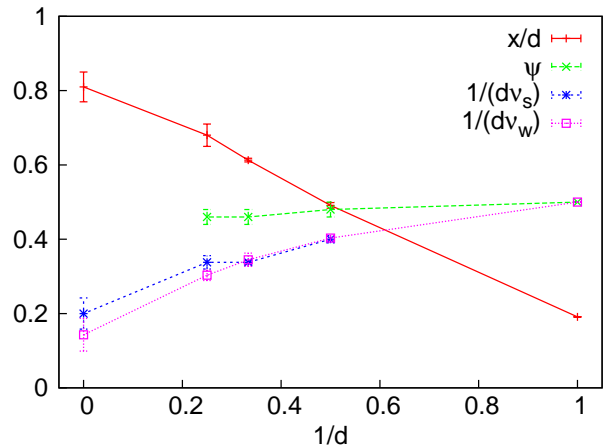}
\end{center}
\vskip -.5cm
\caption{
\label{fig_5} (Color online)
Critical exponents of the RTIM as a function of $1/d$. At $1/d=0$ there are
results of the ER random graph.
}
\end{figure}

As mentioned before the IDFP-s in Table \ref{table:1} control the critical behavior of the spinglass transition, as well as that of
a class of random quantum systems having an order parameter with discrete symmetry, such as random quantum Potts\cite{senthil}
and clock models\cite{carlon-clock-at}. Nonequilibrium phase transitions,
such as the contact process with (strong) disorder\cite{hiv} also belong to this class of universality. The
critical exponents in Table \ref{table:1}, extending with the known results in 1D\cite{fisher} and 2D\cite{2dRG},
show a smooth variation with the dimension, which is presented in Fig.\ref{fig_5}. These results indicate that the
large-$d$ limit of the problem is not singular.
For a given $d$ the correlation length critical exponents, $\nu_s$ and $\nu_w$, agree with each other,
within the error of the calculation\cite{4D}. These satisfy the rigorous bound\cite{ccfs}, $\nu \ge 2/d$, and are in agreement
with the scaling theory at conventional random fixed points\cite{domany,aharony}.
Interestingly, the exponent $\psi$ is found very close to $1/2$ for any considered finite dimension\cite{psi_note}.
This fact can be explained with our observation, that the low-energy excitations in any dimension are
quasi-1D objects and the energy scale can be obtained by renormalizing these objects practically independently of the
rest of the system. This leads to approximately the same type of linear-size dependence of the energy in any dimension.

At this point we want to mention a very recent study by Dimitrova and M\'ezard\cite{dimitrova_mezard} on the critical behavior of
the RTIM by the cavity method\cite{cavity}. In 1D even the simple mean-field cavity method is shown to recover
some of the exact results, such as infinite disorder scaling at the critical point. On the contrary, results on the
Bethe lattice indicate the presence of a conventional random fixed point with a finite dynamical exponent. This
result is probably due to the fact, that the local topology of the Bethe lattice is different from that of the
hypercubic lattices, we considered in this paper. The local topology has already been found to have an
important effect on the critical behavior of random quantum systems\cite{fractal}. For example in the Bethe lattice it seems to be impossible
to define an isotropic and quasi-one-dimensional cluster, which could be relevant for the low-energy excitations and thus
for infinite disorder scaling. A direct SDRG study of the Bethe lattice RTIM could clarify some of the open questions.

Finally we note, that the SDRG investigations presented in this paper can be extended in several directions.
Here we mention the calculation of the entanglement
entropy\cite{refael,lin07,yu07,ladder} in these systems, as well as study of the dynamical
singularities in the disordered and ordered Griffiths phases\cite{im}.

\begin{acknowledgments}
This work has been supported by the Hungarian National Research Fund under grant No OTKA
K62588, K75324 and K77629 and by a German-Hungarian exchange program (DFG-MTA).
We are grateful to D. Huse for helpful correspondence and suggestions and to P. Sz\'epfalusy and H. Rieger for useful discussions.
\end{acknowledgments}


\begin{thebibliography}{99}
\bibitem{2D_performance}
The algorithm described in Ref.\cite{2dRG} has the same performance in 2D, but it is
considerably slower in higher dimensions.
\bibitem{selection} To select site $j$ we first identify a non-decimated end-site to $j$, which is denoted by $k(j)$.
Performing the renormalization around $j$ several nearby sites (including $j$) could be eliminated and $k(j)$ is one
such site, which is not eliminated directly. We measure the effective coupling between $i$ and $k(j)$ as $J_{i,k(j)}$
and its maximal value for different $k(j)$-s is $\tilde{J}_{i,j}$. Now the relevant site to $i$ is such $j$, for which
$\tilde{J}_{i,j}$ is maximal.
\bibitem{constant}
This type of disorder could be relevant for experimental realizations, in which the transverse field is homogeneous,
see c.f. $\rm{LiHo}_x\rm{Y}_{1-x}\rm{F}_4$.
\bibitem{doubling_ER} Between the copies of the ER random graphs we have
drawn $N/2$ random links.
\bibitem{note_distr} The distributions are obtained from histograms, in which in the central part (within the width
of the distribution) there are several thousands of samples. Here the error is smaller than the size of the symbol.
At the tails the error is increasing, but these parts of the distribution have negligible influence on the values
of the calculated critical exponents.
\bibitem{4D}
In 4D the estimates of the two exponents seem to be at the border or slightly outside the estimated error. To
decide about the agreement or disagreement of the two exponents in this case one should study even larger
systems, in particular to reduce the error in the position of the true critical point.
\bibitem{psi_note}
For 1D and 2D we have $\psi=0.5$\cite{fisher} and $\psi=0.48(2)$\cite{2dRG}, respectively.
\end{thebibliography}
\end{document}